\documentclass[a4paper]{jpconf}
\usepackage{graphicx}
\begin{document}
\title{\boldmath{$B_s$} decays at Belle}

\author{Alexey Drutskoy}

\address{Department of Physics, University of Cincinnati, Cincinnati, Ohio, 45221, USA}

\ead{drutskoi@bpost.kek.jp}

\begin{abstract}
We report recent results obtained with the Belle detector
using a 23.6\,fb$^{-1}$ data sample collected
on the $\Upsilon$(5S) resonance
at the KEKB asymmetric energy $e^+ e^-$ collider.
Inclusive semileptonic $B_s^0 \to X^+ l^- \nu$ decays are studied
for the first time and the branching fraction is measured.
Combining the electron and muon channels, we obtain
\mbox{${\cal B}(B_s^0 \rightarrow X^+ l^- \nu)\, =$}
\mbox{$(10.2 \pm 0.8 \pm 0.9)\%$}.
Also, the radiative penguin decay $B_s^0 \to \phi \gamma$ is 
observed for the first time, and an improved upper limit for
the decay $B_s^0 \to \gamma \gamma$ is obtained.
\end{abstract}

\section{Introduction}
Although many $B_s^0$ meson studies have been performed by LEP and 
Tevatron experiments, only a few $B_s^0$ decay branching 
fractions have been measured \cite{pdg}.
Recently an alternative source of $B_s^0$ mesons has 
been explored: $B_s^0$ production by $e^+ e^-$ colliders 
running at the $\Upsilon$(5S) energy. Several important
measurements have been performed recently.

In 2003, the CLEO collaboration collected a data sample 
of 0.42~fb$^{-1}$ at the $\Upsilon$(5S).
Based on this data, in 2005 CLEO published an analysis of
inclusive $D_s$ production at the $\Upsilon$(5S) \cite{cleoi}.
In this paper, the $b\bar{b}$ production cross-section at the $\Upsilon$(5S)
was measured, a procedure to extract the fraction $f_s$ of 
$B_s^{(*)} \bar{B}_s^{(*)}$ events over all $b\bar{b}$ events
% using a large $D_s$ production difference in $B$ and $B_s$ decays
was developed, and the value 
$f_s =$ \mbox{$(16.0 \pm 2.6 \pm 5.8)\%$} was obtained.
In 2006, CLEO published an analysis of exclusive $B_s^0$ decays \cite{cleoe},
where a method, similar to that used at the $\Upsilon$(4S),
was used to extract $B_s^0$ signals. CLEO found
14 candidates consistent with a $B_s^0$ decaying into final states
with a $J/\psi$ or a $D_s$.
These measurements provided a first evidence of $B_s^0$ production at the
$\Upsilon$(5S).

In 2005, the Belle collaboration collected a data sample
of 1.86~fb$^{-1}$ at the $\Upsilon$(5S); this sample is 4.4 times 
larger than the CLEO dataset. Using these data, first $B_s^0$ signals 
with significance larger than 5$\sigma$ were obtained.
In 2007, $D_s$, $D^0$ and $J/\psi$ inclusive production
at the $\Upsilon$(5S) decays was measured \cite{beli} and the 
fraction $f_s =$ \mbox{$(18.0 \pm 1.3 \pm 3.2)\%$} was determinined, 
in good agreement with the CLEO value.
Later, exclusive $B_s$ decays to a $J/\psi$ and a $D_s$ were also 
measured by Belle \cite{bele}, and
the combined $B_s^0$ signal yield obtained was $21\pm\,5$ events. 
With these data, the $B_s^0$ and $B_s^*$ meson masses were also measured.
Our combined $B_s^0$ signal yield disagrees significantly 
with the CLEO measurement (by about factor three).
Moreover, our signal yield is 25$\%$ larger than predicted by
MC simulation, assuming equality of $B^0$ and $B_s^0$
(with $d \to s$ quark replacement) decay branching fractions.
Additionally, we also studied the rare decays
$B_s^0 \to \gamma \gamma$, $B_s^0 \to \phi \gamma$, 
$B_s^0 \to K^+ K^-$ and $B_s^0 \to D_s^{(*)+} D_s^{(*)-}$.
Although no significant signals were
observed with the 1.86~fb$^{-1}$ dataset, background levels in these decays
were accurately estimated and found to be low, indicating good 
potential for future rare $B_s^0$ decay studies
at high luminosity $e^+ e^-$ colliders.

Theoretically almost zero $CP$ violation is expected in $B_s^0$ mixing
within the Standard Model, therefore the search for the $CP$ violation
provides an important opportunity to observe Beyond Standard Model effects.
It is interesting to note that the conventional method of time dependent
$CP$ violation measurement is often assumed to be impossible 
at the $\Upsilon$(5S) due to very fast $B_s^0$ oscillations.
However this is potentially possible, since the corresponding distance 
between oscillation function maximum and minimum at the $\Upsilon$(5S) is 
\mbox{$D = \pi \cdot \Delta m_s \cdot \beta \gamma c = 22.5\,\mu$m}.
Such decay vertex resolution can be reached by existing vertex detectors;
however a beam pipe with a small size or a flat profile should be used.
Another method was proposed in 2004 by the SuperKEKB Physics Working 
Group \cite{sup}: $CP$ violation effects
in $B_s^0$ decays at the $\Upsilon$(5S) can be observed
using decay time information and
the $CP$-even and $CP$-odd $B_s^0$ decay width difference.
This method modifies the general idea proposed 
in Ref. \cite{gros} for high energy hadron colliders. 

In 2006 the Belle collaboration collected at the 
$\Upsilon$(5S) a data sample of
21.7~fb$^{-1}$. This sample provides the possibility to make
the next step in $B_s^0$ studies at the $\Upsilon$(5S):
to move from ``first evidence'' to ``measurements''.
Here, the first results obtained with the
full Belle dataset of 23.6~fb$^{-1}$ are reported.

\section{The branching fraction measurement of semileptonic \boldmath{$B_s^0 \to X^+ l^- \nu$} decay}

Measurements of total semileptonic $B^0$, $B^+$ and $B_s^0$ branching 
fractions, together with corresponding well-measured lifetimes,
provide determination of semileptonic widths.
These widths for the $B^0$, $B^+$ and $B_s^0$ mesons are expected to be
equal, neglecting small corrections due to electromagnetic and
light quark mass difference effects.
The correlated production of a $D_s^+$ meson and a same-sign
lepton at the $\Upsilon$(5S) resonance is used in this analysis 
to measure ${\cal B}(B_s^0 \rightarrow X^+ l^- \nu)$.
The method exploits the fact that, at the $\Upsilon$(5S),
the dominant production of a same-sign $D_s^+$ meson and a fast lepton 
comes from the $B_s^{(*)} \bar{B}_s^{(*)}$ state.
Neither the $c\bar{c}$ continuum nor $B^{(*)}\bar{B}^{(*)}$ states
(except a small contribution due to $\sim 19\%$ $B^0$ mixing effect) 
can result in a 
same-sign $c$-quark (i.e., $D_s$ meson) and primary lepton final state. 
Simultaneous production of a same-sign $D_s^+$ meson and a lepton 
from one $B_s^0$ decay is totally negligible.

To obtain the number of $D_s^+$ mesons produced from
$B_s^0$ decays, we compare the normalized $D_s^+$ momentum
distribution in the $\Upsilon$(5S), $\Upsilon$(4S)
and continuum data samples. The continuum contributions were subtracted 
from the $\Upsilon$(5S) and $\Upsilon$(4S) data samples and the resulting 
distributions are used to calculate the number of $D_s^+$ mesons produced 
from $B_s^0$ decays.

The number of the same-sign $D_s^+$ and lepton events
is obtained at the $\Upsilon$(5S), $\Upsilon$(4S)
and continuum data samples as a function of 
lepton momentum for the whole
$D_s^+$ momentum range $0\ <\ P(D_s^+) \ < 2.6\,$GeV/c.
Backgrounds due to $\Upsilon$(5S)$\,\to B\bar{B}(\pi)$ and
continuum processes, lepton misidentification, photon conversion,
and charged kaon decays are subtracted from the $\Upsilon$(5S)
data to obtain the same-sign $D_s^+$ and lepton events from $B_s^0$ decays.

\begin{figure}[h]
\begin{minipage}{19.6pc}
\includegraphics[width=9.8pc]{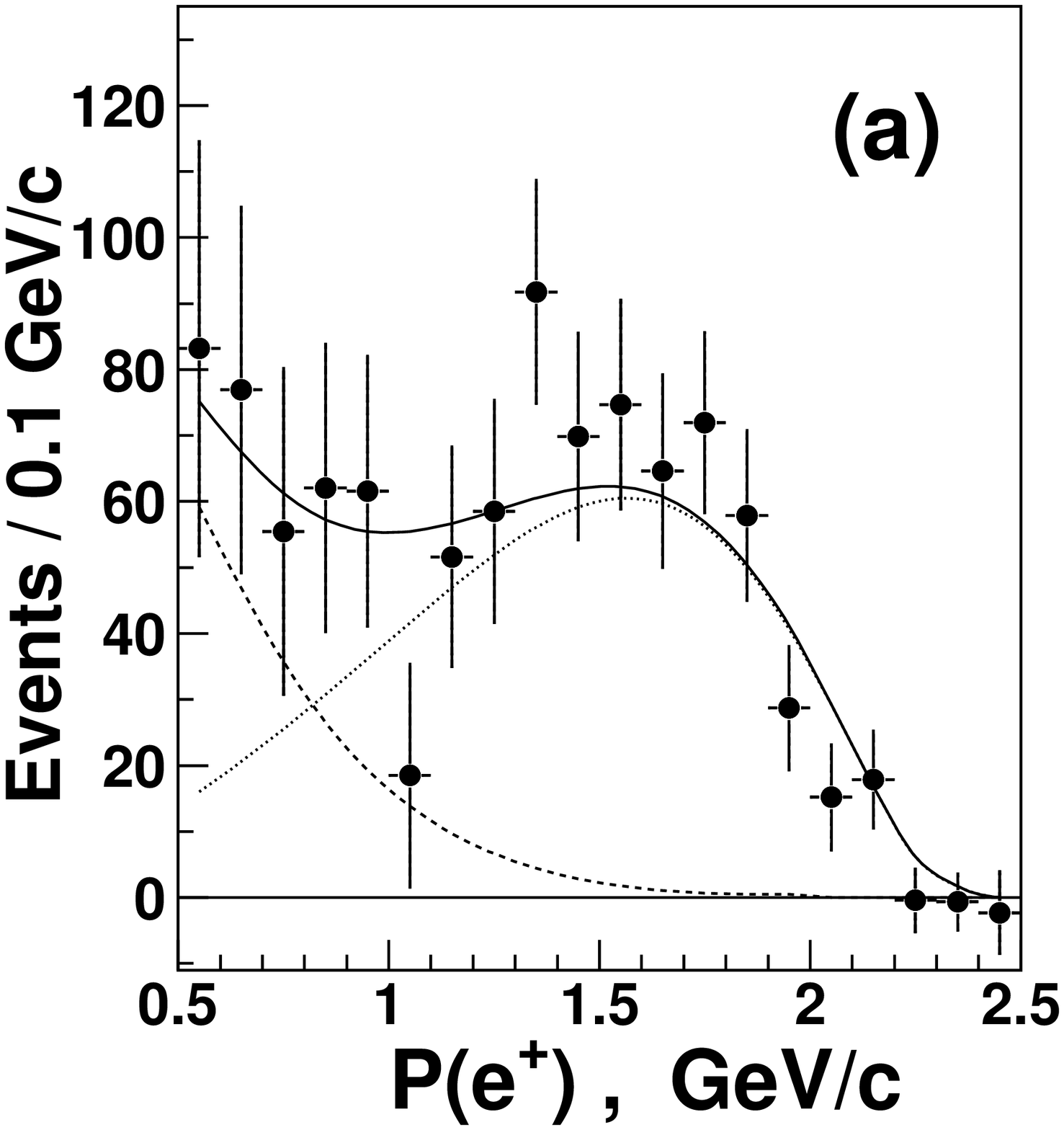}\includegraphics[width=9.8pc]{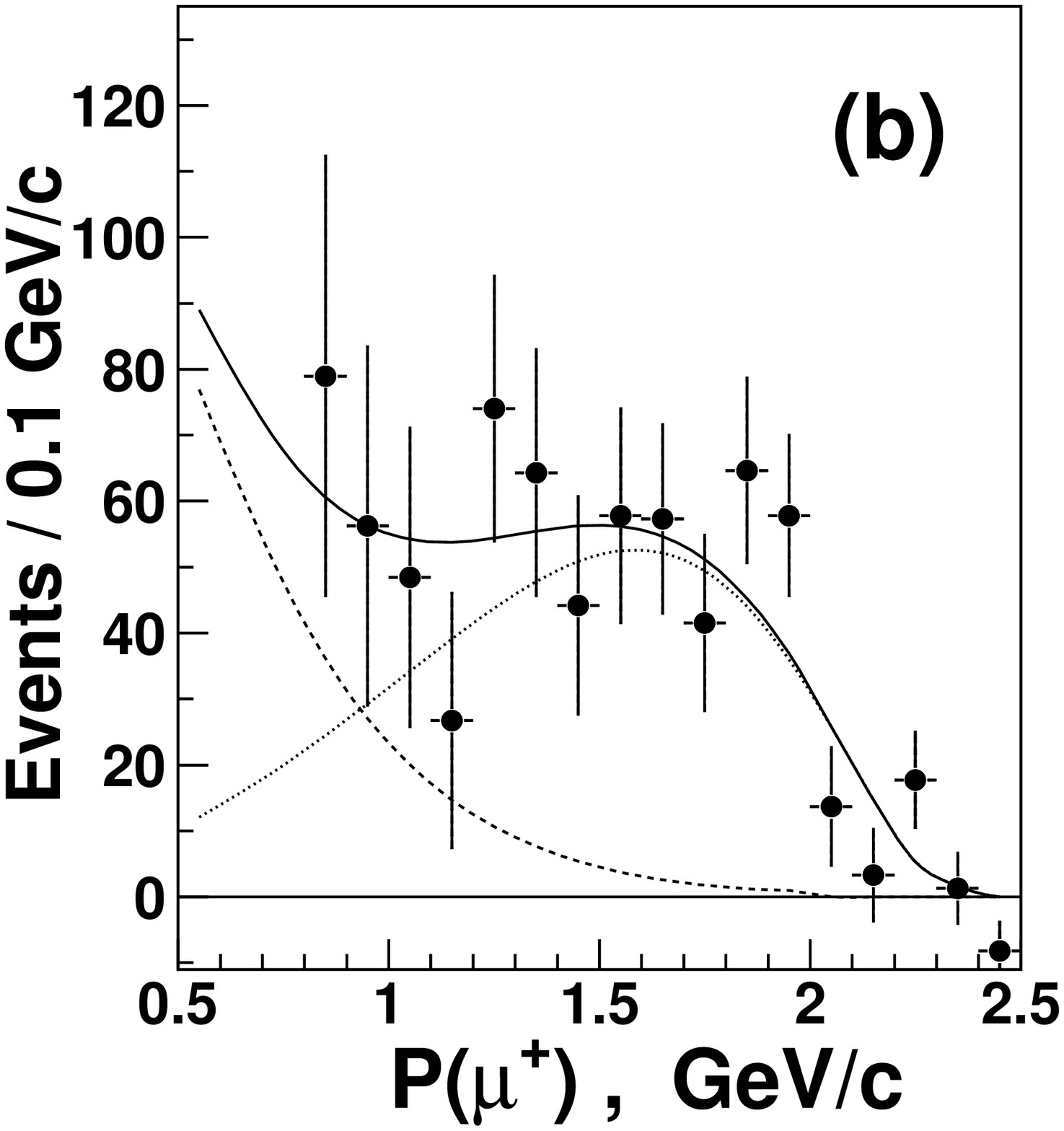}%
\end{minipage}\hspace{2pc}
\begin{minipage}[h]{16pc}\caption{\label{label}The electron (a) and muon (b)
momentum distributions from $B_s^0$ decays. 
The solid curves show the results of the fits, and the dotted 
curves show the fitted contributions from primary and secondary leptons.}
\end{minipage}
\end{figure}

The final lepton momentum distributions produced 
in $B_s^0$ decays
obtained after background subtractions and efficiency correction (Fig. 1)
are used to extract the numbers of primary and secondary leptons.
We fit the data with a function that includes the sum of
two terms with fixed shapes and free normalizations:
one from primary leptons (harder spectrum) and one from secondary leptons.
We fit separately the
electron spectrum and the muon spectrum, as shown in \mbox{Fig.\ 1}. We also
fit the two distributions simultaneously, assuming an equal electron
and muon production rate in $B_s^0$ decays, for both primary and secondary
leptons.

Finally, dividing the numbers of same-sign $D_s^+$ and lepton events
by the full number of $D_s^+$ mesons from $B_s^0$ decays in our $\Upsilon$(5S)
data sample and applying an additional 
factor 1/2 due to the $B_s^0$ mixing effect,
we obtain the following semileptonic branching fractions:

\begin{equation}
{\cal B}(B_s^0 \rightarrow X^+ e^- \nu)\, = (10.9 \pm 1.0 \pm 0.9)\%
\end{equation}
\begin{equation}
{\cal B}(B_s^0 \rightarrow X^+ \mu^- \nu)\, = (9.2 \pm 1.0 \pm 0.8)\%
\end{equation}
\begin{equation}
{\cal B}(B_s^0 \rightarrow X^+ l^- \nu)\, = (10.2 \pm 0.8 \pm 0.9)\% ,
\end{equation}
\noindent
where the last value represents an average over electrons and muons.
The obtained branching fractions are consistent with the PDG value
${\cal B}(B^0 \rightarrow X^+ l^- \nu)\, = (10.33 \pm 0.28)\%$ \cite{pdg},
which is theoretically expected to be approximately the same,
neglecting a small possible lifetime difference and
small corrections due to electromagnetic and
light quark mass difference effects.

\section{The observation of \boldmath{$B_s^0 \to \phi \gamma$} decay and search for \boldmath{$B_s^0 \to \gamma \gamma$} decay} 

The $B_s^0 \to \phi \gamma$ and $B_s^0 \to \gamma \gamma$ are also
studied with the 23.6~fb$^{-1}$ data sample at the $\Upsilon$(5S).
The $B_s^0 \to D_s^- \pi^+$ decay is used to calibrate the central
positions of $M(B_s^*)$ and $\Delta E$ peaks \mbox{(Fig. 2)}. The analysis
of $B_s^0 \to D_s^- \pi^+$ decays is not yet finished, the 
distributions shown are used only for calibration purposes.
Signals are described by Gaussian functions, whereas backgrounds are
described by a linear function in $\Delta E$ and a so-called ARGUS function
in $M_{\rm bc}$. The data corresponding to 
the $B^0_s\bar{B}^0_s$, $B^0_s\bar{B}^*_s$, $B^*_s\bar{B}^0_s$,
and $B^*_s\bar{B}^*_s$ channels are combined.

\begin{figure}[h]
\begin{minipage}{27pc}
\includegraphics[width=13.5pc]{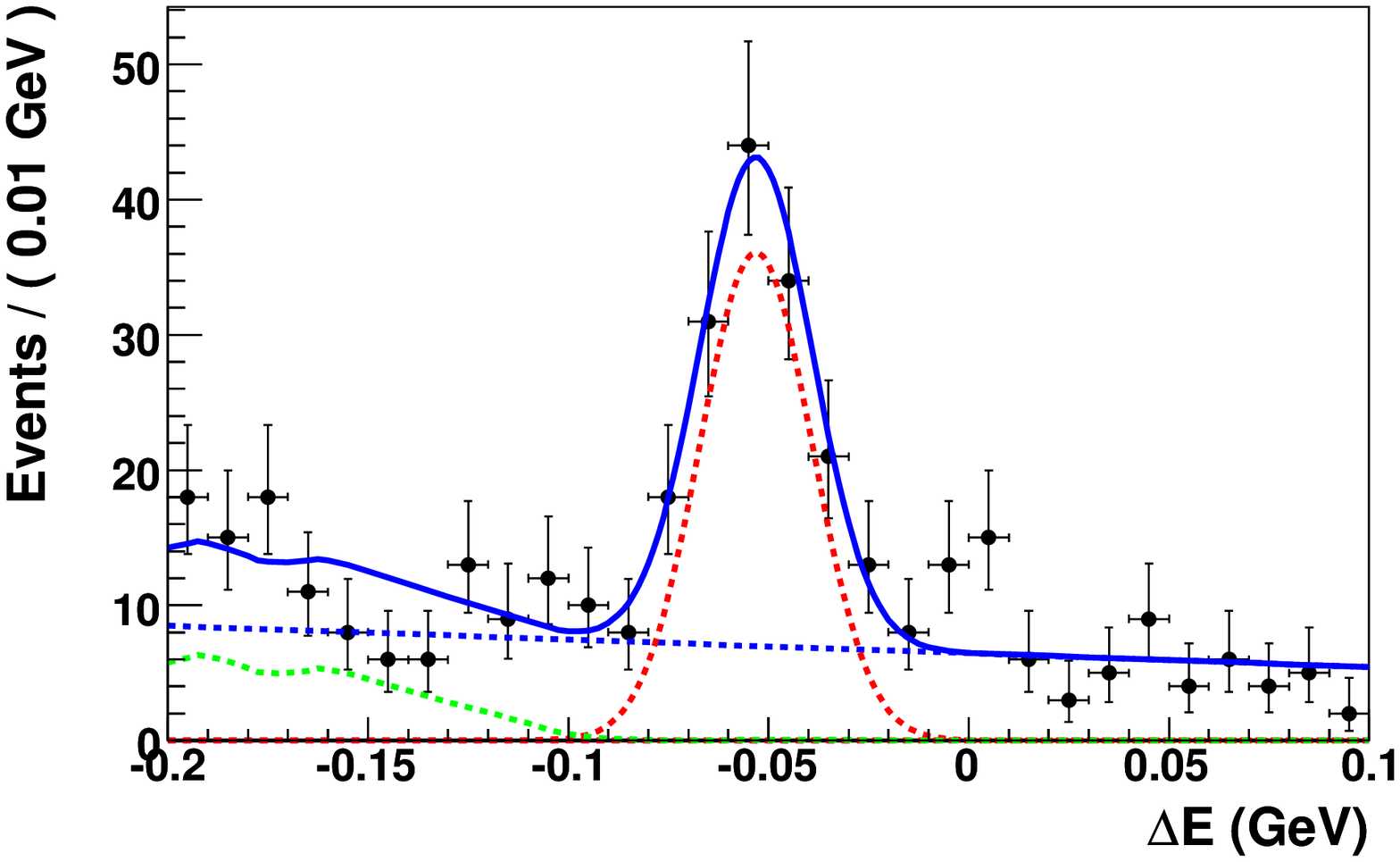}\includegraphics[width=13.5pc]{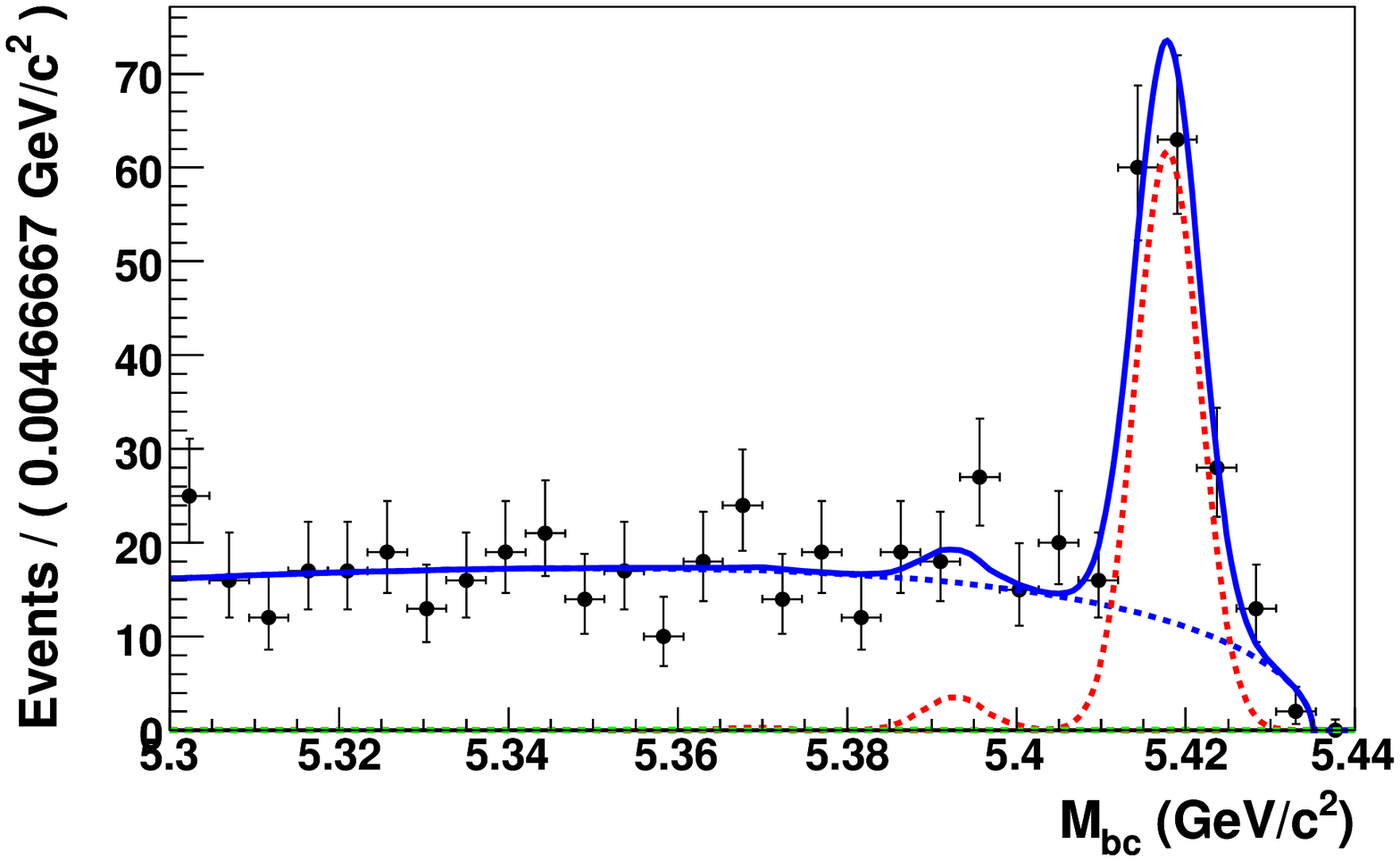}%
\end{minipage}\hspace{2pc}
\begin{minipage}[h]{8.pc}\caption{\label{label}The $\Delta E$ (left) and 
$M_{\rm bc}$ (right) distributions for the $B_s^0 \to D_s^- \pi^+$ decay.}
\end{minipage}
\end{figure}

Finally we have measured the branching fraction 
${\cal B}(B_s^0 \to \phi \gamma) = 
(5.7^{+1.8}_{-1.5}{\rm (stat)}^{+1.2}_{-1.1}{\rm (syst)})\times 10^{-5}$
and obtained an upper limit at the 90$\%$ confidence level of 
${\cal B}(B_s^0 \to \gamma \gamma) < 8.7\times 10^{-6}$.
These analyses are described in detail in the talk of J. Wicht
at this conference.

\end{document}